\documentclass[fleqn,twoside]{article}
\usepackage{espcrc2}
\usepackage{graphicx}
\usepackage[figuresright]{rotating}

\newcommand{\no}{\nonumber}

\newcommand{\simg}{\rlap{\raise -4pt \hbox{$\sim$}}
                   \raise 3pt \hbox{$>$}}
\newcommand{\siml}{\rlap{\raise -4pt \hbox{$\sim$}}
                   \raise 3pt \hbox{$<$}}

\newcommand{\lo}{{(0)}}
\newcommand{\latt}{{\rm latt}}

\newcommand{\msbar}{{\overline {\rm MS}}}

\def\ovec{\partial_\mu\hspace{-0.4cm}\raisebox{1.8ex}{$\rightarrow$}}
\def\antivec{\partial_\mu\hspace{-0.4cm}\raisebox{1.8ex}{$\leftarrow$}}

\newcommand{\mph}{{m_{Q}}}
\newcommand{\mphlo}{\mph^{(0)}}

\title{
 One-loop determination of mass dependent $O(a)$ improvement
 coefficients for the heavy-light vector and axial-vector currents with
 relativistic heavy and domain-wall light quarks
 \thanks{Talk presented by N. Yamada.}
 \thanks{
  The calculation of this work were mainly done on Riken Super Combined
 Cluster System at Riken.
         }
}
\author{Norikazu Yamada\address[RBRC]{
         RIKEN BNL Research Center, Brookhaven National Laboratory,
         Upton, NY 11973, USA}, 
        Sinya Aoki$^{\rm a,b}$ and
        Yoshinobu Kuramashi\address[Tsukuba]{
         Institute of Physics, University of Tsukuba, 
         Tsukuba, Ibaraki 305-8571, Japan}
}

\begin{document}
\begin{abstract}
 We present the one-loop results of the mass dependent $O(a)$
 improvement coefficients for the heavy-light vector and axial-vector
 currents consisting of the relativistic heavy and the domain-wall light
 quarks.
 The calculations are performed with the plaquette, Iwasaki and DBW2
 gauge actions.
 The heavy quark mass and domain-wall height dependence is
 investigated.
 We point out that the exact chiral symmetry held by the lattice light
 quark action leads to an exact relation between the improvement
 coefficients for the vector and axial-vector currents without regard to
 the lattice heavy quark action.
\vspace{-2ex}
\end{abstract}

\maketitle

 \section{Introduction}
 \label{sec:intro}

 Lattice QCD has played an essential role in the nonperturbative study
 of the heavy-light physics~\cite{Wingate:2004},
 and further progress, especially in precision, is expected.
 In the lattice study of heavy-light physics, the scaling violations due
 to the large heavy quark mass and the uncertainty arising from long
 extrapolations to the physical light quark mass have been significant
 sources of uncertainty, and toward more precision studies it is
 indispensable to reduce these systematic errors.
 Combining the relativistic heavy quark~\cite{Aoki:2001ra,Aoki:2003dg}
 and the domain-wall (DW) light
 quark~\cite{Kaplan:1992bt,Shamir:1993zy,Furman:ky}
 is a promising choice to simulate heavy-light system.
 In the following, we present the one-loop calculation of the $O(a)$
 improvement coefficients for the heavy-light 
 current consisting of the these quarks.
 The detailed descriptions are available in Refs.~\cite{Yamada:2004ri}.

 \section{Quark actions}
 \label{sec:actions}

  The relativistic heavy quark action~\cite{Aoki:2001ra},
 \begin{eqnarray}
&&\hspace*{-4ex}  S_Q
= \sum_x
  {\bar Q}(x)\bigg[\  m_0 +\gamma_0 D_0 + \nu_Q \sum_i\gamma_i D_i\no\\
      && - \frac{r_t a}{2}D_0^2
         - \frac{r_s a}{2} \sum_i D_i^2
         - \frac{ig a}{2}c_E \sum_i\sigma_{0i}F_{0i}
\no\\&&
         - \frac{ig a}{4}c_B \sum_{i,j}\sigma_{ij}F_{ij}\
  \bigg]Q(x),
\label{eq:action_clover}
 \end{eqnarray}
 is obtained by pushing the on-shell improvement
 program~\cite{Luscher:1984xn} to the massive Wilson fermions, and is
 equal to the Fermilab action~\cite{El-Khadra:1996mp} with a special
 choice of parameters.
 The four parameters, $\nu_Q,r_s,c_E,c_B$,
 have been known to the one-loop level~\cite{Aoki:2003dg}.
 
 In this work, we employ the following version of the DW action,
 \begin{eqnarray} 
 &&\hspace{-3ex}
    S_{\rm dw}
  = \sum_{x,y}\sum_{s,t}
    \bar{\psi}_s(x)D_{\rm dw}(x,s;y,t)\psi_t(y)\no\\
 &&\hspace{6ex}
  - \frac{R_s}{2}\ \bar{q}(x)\ \sum_iD_i^2\ q(x), 
\label{eq:action_dw}\\
 &&\hspace{-3ex}
    D_{\rm dw}(x,s;y,t)
  = D_{\rm w}(x,y)\,\delta_{s,t}
  + \delta_{x,y}D_5(s,t),\\
 &&\hspace{-3ex}
    D_{\rm w}(x,y)
  =  \gamma_0 D_0 + \nu_q \sum_i\gamma_iD_i\no\\
 &&\hspace{8ex}
  - \frac{a}{2} \sum_\mu D_\mu^2
  - M_5\,\delta_{x,y},
\label{eq:wilson-dirac-op}\\
 &&\hspace{-3ex}
    D_5(s,t)
  = \delta_{s,t} - P_L\,\delta_{s+1,t} - P_R\,\delta_{s-1,t}\no\\
 &&\hspace{6ex}
     + m_f\left[   P_L\,\delta_{s,N_5}\delta_{1,t}
                 + P_R\,\delta_{s,1}  \delta_{N_5,t}
          \right],
 \end{eqnarray}
 where $P_{R/L}=(1\pm\gamma_5)/2$, and $s$ and $t$ label the coordinate
 of the fifth dimension running 1 to $N_5$ ($\rightarrow\infty$ in this
 work).
 The physical quark field $q(x)$ is defined by
 \begin{eqnarray}
      q(x)
  &=& P_L\psi_1(x) + P_R\psi_{N_5}(x),\\
      \bar{q}(x)
  &=& \bar{\psi}_1(x)P_R + \bar{\psi}_{N_5}(x)P_L.
 \end{eqnarray}
 In the action, we have introduced two new parameters $\nu_q$ and $R_s$
 according to the following reason.
 It was found that some of the improvement coefficients for the
 heavy-light currents can be determined only if the light quark mass is
 finite~\cite{Aoki:2004th}, therefore we need to consider the case that
 $m_f$ is finite.
 However, once the quark acquires a finite mass, $m_f a$ corrections to
 the dispersion relation causes unwanted infrared
 divergences in the loop integrations.
 The new parameters $\nu_q$ and $R_s$, where $\nu_q\rightarrow 1$ and
 $R_s\rightarrow 0$ in the massless limit, are to adjust the dispersion
 relation and eliminate the infrared divergences.
 Perturbative technique for the DW quarks is described in
 Ref.~\cite{Aoki:1998vv}.
 It should be noted that the both actions restore the exact space-time
 rotational symmetry in the massless limit.

 \section{Properties of the coefficients}
 \label{sec:definition-imp-coef}

 We define the improvement coefficients, $\Delta_{V_\mu}$,
 $c_{V_\mu^{\pm,HL}}$, for the vector current as follows,
 \begin{eqnarray}
 &&\hspace*{-3ex}
     V^{\msbar}_\mu
 =   Z_{V_\mu}^{\msbar-\latt}\,V^{\rm lat,imp}_\mu,\no\\
 &&\hspace*{-3ex}
     Z_{V_\mu}^{\msbar-\latt}
 =   Z_{\rm wf}\,\left[ 1 - g^2 \Delta_{V_\mu} \right],\no\\
 &&\hspace*{-3ex}
     V^{\rm lat,imp}_\mu
 =   {\bar q} \gamma_\mu Q
 - g^2 c_{V_\mu}^+\partial_\mu^- \{{\bar q} Q\}
 \no\\ &&\hspace*{5.5ex}
 -\, g^2 c_{V_\mu}^- \partial_\mu^+ \{{\bar q} Q\}
 - g^2 c_{V_\mu}^L \{{\vec\partial_i}{\bar q}\}
                      \gamma_i\gamma_\mu Q
 \no\\ &&\hspace*{5.5ex}
 -\, g^2 c_{V_\mu}^H {\bar q}\gamma_\mu \gamma_i
                      \{{\vec \partial_i} Q\} + O(g^4),
 \label{eq:v_r}
 \end{eqnarray}
 where $\partial^\pm$=$\ovec\pm\antivec$, $q$ and $Q$ are lattice light
 and heavy quark fields, respectively, and $Z_{\rm wf}$ is a known
 factor relating the wave function renormalizations for both quarks.
 While, in general, the improvement coefficients for the temporal and
 spacial components have different values, in the massless limit they agree
 because of the restoration of space-time rotational symmetry.
 The equation of motion always allows us to set $c^H_{V_0}=c^L_{V_0}=0$.

 The exact chiral symmetry of the DW fermions brings an exact
 relation between the improvement coefficients for the vector and
 axial-vector currents (for the proof, see Ref.~\cite{Yamada:2004ri}).
 Thanks to this relation, once  $Z_{V_\mu}^{\msbar-\latt}$ and
 $c_{V_\mu}^{\pm,H,L}$ are known, with these coefficients the continuum
 (improved) axial-vector current is immediately given by
 \begin{eqnarray}
 &&\hspace*{-4ex}
      A^{\msbar}_\mu
  =   Z_{V_\mu}^{\msbar-\latt}\,A^{\rm lat,imp}_\mu,\no\\
 &&\hspace*{-4ex}
      A^{\rm lat,imp}_\mu
  =   {\bar q} \gamma_\mu\gamma_5 Q
    + g^2 c_{V_\mu}^-\partial_\mu^+\{{\bar q}\gamma_5 Q\}\no\\
 &&\hspace*{3ex}
    +\, g^2 c_{V_\mu}^+\partial_\mu^-\{{\bar q}\gamma_5 Q\}
    + g^2 c_{V_\mu}^L\{{\vec\partial_i}{\bar q}\}
                     \gamma_i \gamma_\mu\gamma_5 Q\no\\
 &&\hspace*{3ex}
    -\, g^2 c_{V_\mu}^H {\bar q}\gamma_\mu\gamma_5 \gamma_i
                      \{{\vec \partial_i} Q\}+O(g^4).
 \label{eq:a_r}
 \end{eqnarray}
 This property holds irrelevantly to heavy quark action, and this
 statement can be extended to any pair of operators which belong to the
 same chiral multiplet.

 The relation $Z_{V_\mu}^{\msbar-\latt}$=$Z_{A_\mu}^{\msbar-\latt}$ is
 especially useful for the test of the soft pion relation
 $f^0(q^2_{\rm max})$=$f_B/f_\pi$~\cite{Kitazawa:1993bk},
 because the unknown two loop coefficient in the ratio has prevented the
 test from being more precise.

 \section{Results}
 \label{sec:determination}

 The loop integrations on the lattice are performed by a mode sum
 and BASES. 
 The dependence of the coefficients on the tree level pole mass of heavy
 quark, $\mph^\lo$, and the domain-wall height, $M_5$, are searched by
 choosing several values for both parameters.
 We checked that the two methods give consistent values if
 $m_Q^\lo < 7.0$ and $0.1<M_5<1.9$, and found that in that region the
 discrepancy is less than 0.001, which can only affect physical
 quantities by 0.5\%, at most.
 The continuum part is evaluated in the naive dimensional regularization
 (NDR) with the modified minimal subtraction scheme ($\msbar$).

\begin{figure}[t]
  \vspace*{-0ex}
  \begin{center}
   \includegraphics*[width=7.3cm,clip=true]{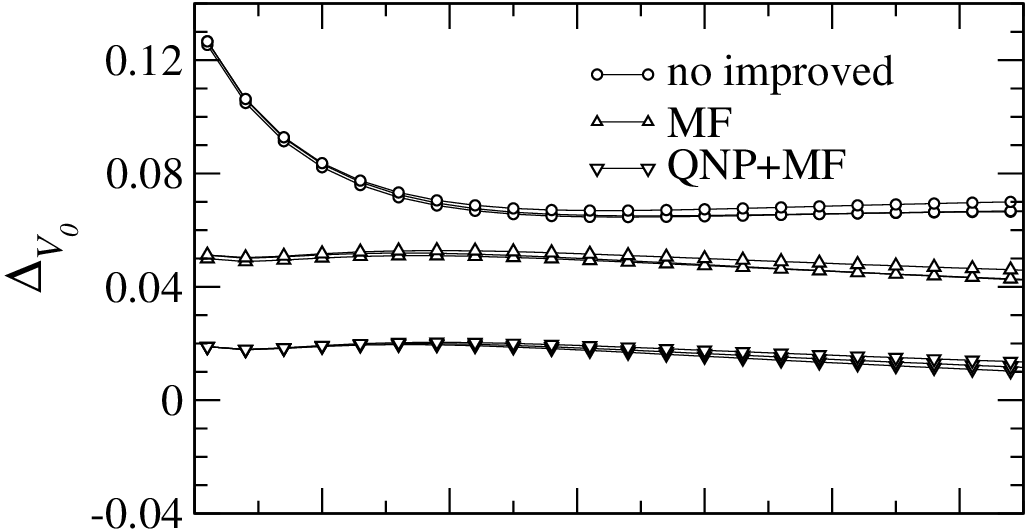}\\
   \includegraphics*[width=7.3cm,clip=true]{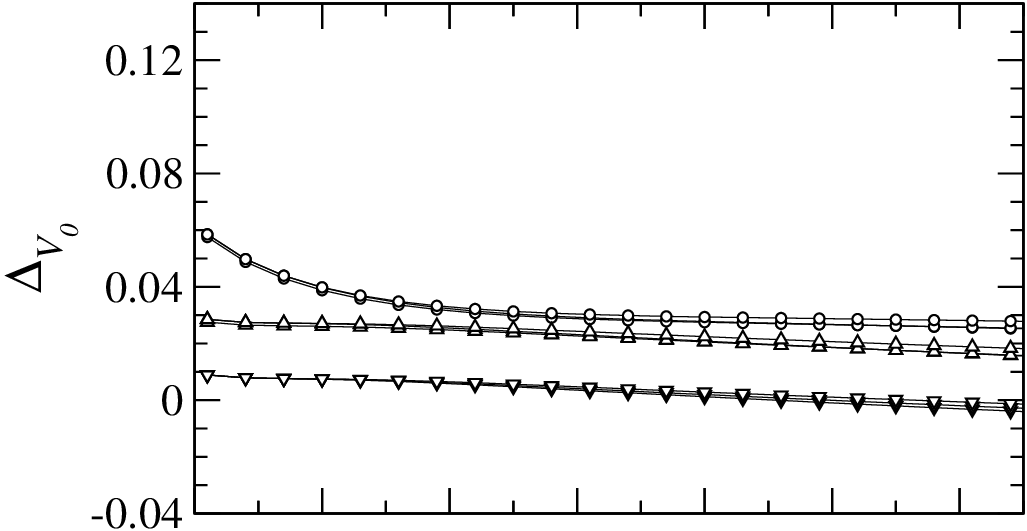}\\
   \includegraphics*[width=7.3cm,clip=true]{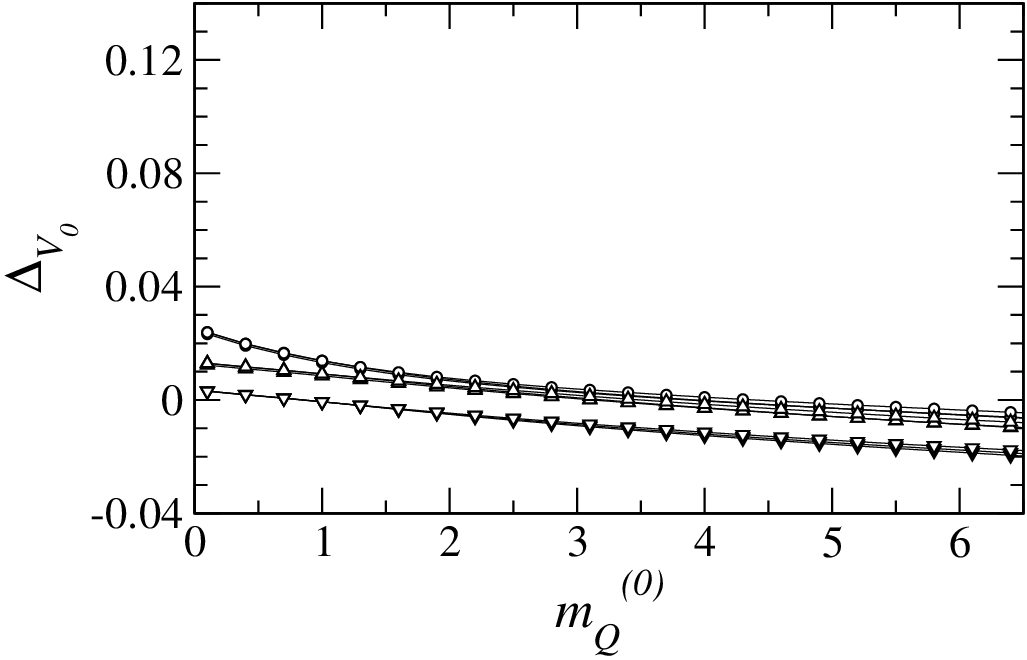}
   \vspace{-7ex}
  \end{center}
  \caption{The $m_Q^\lo$ dependence of $\Delta_{V_{0}}$,
  $\Delta_{V_{0}}^{\rm MF}$ and $\Delta_{V_{0}}^{\rm QNP+MF}$ with the
  plaquette, Iwasaki and DBW2 gauge action from top to bottom.
  }
  \label{fig:zvk0}
  \vspace{-4ex}
 \end{figure}

 In Fig.~\ref{fig:zvk0}, the $\mphlo$ dependence of $\Delta_{V_0}$
 (circles) with three different gauge actions, the plaquette,
 Iwasaki~\cite{Iwasaki:1983ck} and DBW2~\cite{Takaishi:1996xj}, is shown
 as an example.
 The results with $M_5$=0.5, 1.1 and 1.7 are shown as examples to give
 some idea about the $M_5$ dependence, which is turned out to be smaller
 than the $m_Q$ dependence in the most cases.
 The solid lines are obtained by fitting the numerical data to a certain
 functional form.
 The fit reproduces the data very well over the whole region of
 $m_Q^\lo$ and $M_5$ as seen in the figures.

 \section{Improving perturbation series}
 \label{sec:imp-of-ptseries}

 We examined two ways to improve the perturbation series.
 One is the mean field improvement (MF)~\cite{Lepage:1992xa}
 using the mean plaquette value, and another is involving the
 nonperturbative renormalization factor for the light-light vector
 current consisting of the DW fermions in addition to the mean
 field improvement.
 The latter which we call QNP+MF has an advantage because the wave
 function renormalization for the DW fermion is taken care of
 in all orders.

 Reorganizing ${\Delta}_{V_0}$ with two ways, we worked out two
 coefficients, $\Delta_{V_{0}}^{\rm MF}$ (triangle-up) and
 $\Delta_{V_{0}}^{\rm QNP+MF}$ (triangle-down), which are plotted in
 Fig.~\ref{fig:zvk0} for the three different gauge actions.
 From the figure, it is seen that in the non-improved case the size of
 the one-loop coefficient largely depends on the gauge action while once
 one introduced the quasi non-perturbative and mean field improvements
 together the dependence becomes less transparent.
 It is also interesting to see that the mean field improvement makes the
 $m_Q$ dependence smaller.
 The similar findings are made for the improvements of $\Delta_{V_k}$ as
 well.

\vspace{2ex}
 This work is supported in part by the Grants-in-Aid for
 Scientific Research from the Ministry of Education, 
 Culture, Sports, Science and Technology.
 (Nos. 13135204, 15204015, 15540251, 15740165, 16028201.)

\bibliography{basename of .bib file}

\end{document}